\documentclass[letter]{aa} 
\usepackage{graphicx}
\usepackage{txfonts}
\usepackage{xcolor}
\begin{document}

   \title{Gas vs. solid phase deuterated chemistry: HDCO and D$_2$CO in massive star-forming regions}

   \author{S. Zahorecz
          \inst{1,2,3,4,5}
          \and
          I. Jimenez-Serra
          \inst{6}
          \and
          L. Testi
          \inst{4,7,9} 
          \and
          K. Immer
          \inst{4}
		  \and
          F. Fontani
          \inst{7}
		  \and
          P. Caselli
          \inst{8}
	      \and
          K. Wang
          \inst{4}
          \and
          L. V. Toth
          \inst{5}
          }

   \institute{Department of Physical Science, Graduate School of Science, Osaka Prefecture University, 1-1 Gakuen-cho, Naka-ku, Sakai, Osaka 599-8531, Japan
	  \email{s.zahorecz@p.s.osakafu-u.ac.jp}
			\and
		National Astronomical Observatory of Japan, National Institutes of Natural Science, 2-21-1 Osawa, Mitaka, Tokyo 181-8588, Japan    
			\and   
		Konkoly Observatory, Research Centre for Astronomy and Earth Sciences, Hungarian Academy of Sciences, 1121 Budapest, Konkoly Thege Mikl\'os \'ut 15-17, Hungary,
            \and
        European Southern Observatory, Karl-Schwarzschild-Str. 2, 85748, Garching bei M\"unchen, Germany
         \and
         	E\"otv\"os Lor\'and University, Department of Astronomy, P\'azm\'any P\'eter s\'et\'any 1/A, 1117, Budapest, Hungary
         \and
         School of Physics and Astronomy, Queen Mary University of London,
Mile End Road, E1 4NS London, United Kingdom
		\and
        INAF-Osservatorio Astrofisico di Arcetri, L.go E. Fermi 5, 50125 Firenze, Italy
 		\and
        Max-Planck-Institut f\"ur Extraterrestrische Physik, Giessenbachstrasse 1, 85748 Garching bei M\"unchen, Germany
		\and
               Excellence Cluster Universe, Boltzmannstr. 2, 85748, Garching bei M\"unchen, Germany
             }

   \date{Received ; accepted }

 
  \abstract
  {The formation of deuterated molecules is favoured at low temperatures and high densities. Therefore, the deuteration fraction (D$_{frac}$) is expected to be enhanced in cold, dense prestellar cores and to decrease after protostellar birth. Previous studies have shown that the deuterated forms of species such as N$_2$H$^+$ (formed in the gas phase) and CH$_3$OH (formed on grain surfaces) can be used as evolutionary indicators and to constrain their dominant formation processes and time-scales.}
   {Formaldehyde (H$_2$CO) and its deuterated forms can be produced both in the gas phase and on grain surfaces. However, the relative importance of these two chemical pathways is unclear. Comparison of the deuteration fraction of H$_2$CO with respect to that of N$_2$H$^+$, NH$_3$ and CH$_3$OH can help us to understand its formation processes and time-scales.}
   {
  With the new SEPIA Band 5 receiver on APEX, we have observed the $J$=3$\rightarrow$2 rotational lines of HDCO and D$_2$CO at 193 GHz and 175 GHz toward three massive star forming regions hosting objects at different evolutionary stages: two High-mass Starless Cores (HMSC), two High-mass Protostellar Objects (HMPOs), and one Ultracompact HII region (UC~HII). By using previously obtained H$_2$CO $J$=3$\rightarrow$2 data, the deuteration fractions HDCO/H$_2$CO and D$_2$CO/HDCO are estimated.
   }
   {
   Our observations show that singly-deuterated H$_2$CO is detected toward all sources and that the deuteration fraction of H$_2$CO increases from the HMSC to the HMPO phase and then sharply decreases in the latest evolutionary stage (UCHII). The doubly-deuterated form of H$_2$CO is detected only in the earlier evolutionary stages with D$_2$CO/H$_2$CO showing a pattern that is qualitatively consistent with that of HDCO/H$_2$CO, within current uncertainties. 
   }
   {
   Our initial results show that H$_2$CO may display a similar D$_{frac}$ pattern as that of CH$_3$OH in massive young stellar objects. This finding suggests that solid state reactions dominate its formation.
   }

   \keywords{ISM: molecules; radio lines: ISM; stars:formation; Astrochemistry}

   \maketitle
%

\section{Introduction}
In molecular cloud cores, the formation of deuterated molecules is favoured at low temperatures (T $\leq$ 20~K) and at high densities (n $\geq 10^4$~cm$^{-3}$). It is thus expected that the deuteration fraction (D$_{frac}$, the relative abundance between a species containing D as compared to the same species containing H) 
is enhanced in cold and dense prestellar cores.  
D$_{frac}$ should then decrease after protostellar birth, when the young stellar object heats up the central region of the core \citep{2002P&SS...50.1133C}.  For low mass star forming cores, observations of deuterated species produced in gas-phase reactions, such as H$_2$D$^+$ and N$_2$D$^+$, have confirmed this theoretical scenario \citep[][]{2005ApJ...619..379C, 2008A&A...492..703C}. 

Recent studies show that high D$_{frac}$ values are also typical for high mass star forming cores and that the D$_{frac}$ of some species could be an evolutionary indicator also in the intermediate and high mass regime \citep[e.g. ][]{2010A&A...517L...6B, 2011A&A...529L...7F, 2012ApJ...747..140S}.
High-mass star forming regions can be divided into the following evolutionary stages: high-mass starless cores (HMSCs), high-mass protostellar objects (HMPOs) and ultracompact HII regions (UCHIIs) \citep[e.g. ][]{2007prpl.conf..165B,2014prpl.conf..149T}.
By studying several deuterated species in 27 massive cores, \citet{2011A&A...529L...7F,2015A&A...575A..87F} found that species formed exclusively in the gas (N$_2$H$^+$) showed different evolutionary trends from those formed partially (NH$_3$) or totally (CH$_3$OH) on grain mantles. Indeed, the abundance of N$_2$D$^+$ is higher in HMSCs and it drops by about an order of magnitude during the HMPO and UCHII stages. This is due to the higher gas temperatures found in the latter objects that allow the destruction of H$_2$D$^+$ via the endothermic reaction H$_2$D$^+$ + H$_2$ $\rightarrow$ H$_3^+$ + HD \citep[see][]{2002P&SS...50.1275G, 2011A&A...529L...7F}. In contrast, deuterated methanol, formed only on grain surfaces, is detected towards HMPOs and externally heated HMSCs only, possibly as a result of evaporation/sputtering of grain mantles. Therefore, while D$_{frac}$(N$_2$H$^+$) can be used as an indicator of the initial conditions in starless/pre-stellar cores, high values of D$_{frac}$(CH$_3$OH) are a good probe of the earliest protostellar phases \citep{2015A&A...575A..87F}. The deuteration fraction of NH$_3$, whose formation pathways occur both in the gas-phase and on grain surfaces, does not show statistically significant changes with evolution and thus, there is not a dominant formation pathway for NH$_3$ and its deuterated forms.

Like NH$_3$, formaldehyde (H$_2$CO) may also be produced both in the gas phase and on grain surfaces. The two main pathways for the production of H$_2$CO (and its deuterated counterparts) involve CH$_3^+$ (viz. CH$_2$D$^+$ and CHD$_2^+$) in the gas phase and multiple hydrogenation (viz. deuteration) of CO in the ices \citep[see e.g.][]{2007A&A...471..849R}. The gas phase pathway is similar to the one of N$_2$H$^+$ with the main difference that deuteration of H$_2$CO can also happen at warmer temperatures \citep[30-50\,K, see][]{2009A&A...508..737P}, whereas T $<$ 20\,K are needed to increase the deuteration of N$_2$H$^+$ via H$_2$D$^+$. The ice phase formation route of H$_2$CO is similar to that of CH$_3$OH and their deuterated forms. Although laboratory work and observational studies of low-mass protostars suggest an important contribution from grain surface chemistry for the production of H$_2$CO \citep[][]{2005IAUS..231..415W,2007A&A...471..849R, 2011A&A...527A..39B}, the relative importance of the dust grain vs gas-phase formation routes remains unclear. Furthermore, D$_2$CO measurements in intermediate- and in high-mass star-forming regions are lacking \citep[D$_2$CO has been firmly detected toward NGC 7129-FIRS 2 and tentatively toward the MonR2 ultra-compact HII region;][]{2005A&A...444..481F,2014A&A...569A..19T}, and it is thus unknown whether the chemistry of H$_2$CO and of its deuterated counterparts is governed by the same mechanisms in both low-mass and more massive objects.

In this Letter, we report the first detection of doubly-deuterated formaldehyde, D$_2$CO, toward high-mass star-forming cores. These are the initial findings of a search for HDCO and D$_2$CO emission toward three high-mass star-forming regions at different evolutionary stages using the new SEPIA Band 5 receiver available at the Atacama Pathfinder EXperiment (APEX\footnote{This publication is based on data acquired with the Atacama Pathfinder Experiment (APEX). APEX is a collaboration between the Max-Planck-Institut f\"ur Radioastronomie, the European Southern Observatory, and the Onsala Space Observatory.}) telescope. The targets were extracted from the sample of \citet{2011A&A...529L...7F} for which the deuteration fraction of other molecules (e.g. N$_2$H$^+$, CH$_3$OH and NH$_3$) has been measured \citep{2015A&A...575A..87F}. The evolutionary stages of our sample go from the initial conditions in HMSCs, to the HMPOs phase and the UCHII regions. The source sample and the observations are described in Section 2, while our results are presented in Section 3. In Section 4, we discuss our results and put them in context with respect to previous findings in low-mass star-forming regions. In Section 5, we summarize our conclusions.

\section{Observations}
\subsection{Selected targets}
The selected sources are extracted from \citet{2011A&A...529L...7F} and are: AFGL5142, IRAS 05358+3543 and G5.89-0.39. These are the sources with the brightest lines of H$_2$CO within each evolutionary stage. We note that this selection may introduce biases, therefore a follow-up study with larger sample is needed.
The bolometric luminosities of the objects are 10$^{3.6}$, 10$^{3.8}$ and 10$^{5.1}$ L$_\odot$ and the integrated gas masses are 210, 300 and 300 M$_\odot$ for AFGL5142, IRAS05358 and G5.89, respectively \citep{2011A&A...529L...7F, 2016ApJ...824...31L,2002ApJ...566..945B,2005ApJ...633..535B,2009ApJ...695.1399T}. AFGL5142 is a high-mass star-forming region (distance of 2.14$\pm{0.05}$\,kpc; Burns et al. in prep.), which hosts several dense cores at different evolutionary stages: the central core (CC, classified as HMPO), the western core (WC, a HMSC core), and the eastern core \citep[EC; see][]{2011A&A...525A.141B,2011A&A...529L...7F}.
We centered our single-pointing observations on CC. However, since the separation between WC and CC is 9$"$, and since the APEX beam is $\sim$34$\arcsec$ at 190$\,$GHz, WC and CC were also covered in our observations. 
The N$_2$H$^+$ (1-0) emission in AFGL5142 shows two velocity components at $-$2 and $-$4 km s$^{-1}$, associated respectively with CC and WC \citep{2011A&A...525A.141B}. 

IRAS 05358+3543, located at 1.8 kpc \citep[][]{snell1990}, contains three dust condensations (mm1, mm2, mm3) within an area of 9$\arcsec\times$4$\arcsec$  \citep{beuther2002}. According to \citet{2007A&A...475..925L}, mm1 splits into mm1a, a hot core and a massive circumstellar disk with T$\sim$220K, and mm1b, which is at an earlier stage of evolution. Source mm2 is a low-to-intermediate mass protostar while mm3 is a HMSC. Our observation is centered on mm1, but all three dust condensations are covered within the APEX beam.

G5.89-0.39 is a shell-like UCHII region (diameter of $\sim$4$\arcsec$) found at 2.99$^{+0.19}_{-0.17}$\,kpc \citep{sato2014}. Sub-arcsecond observations reveal at least five dust condensations \citep{1538-4357-704-1-L5}.  We centered our observations on the SMA-N dust condensation. Several energetic outflows and maser activity have been detected toward G5.89-0.39 \citep[see e.g. ][]{2008ApJ...680.1271H, 2005ApJS..160..220F}.

\subsection{IRAM-30m H$_2$CO observations}

Previous observations of these objects were performed using the IRAM-30m telescope \citep[details can be found in][]{2011A&A...529L...7F,2015A&A...575A..87F}. The frequency setups included the 3$_{03}$-2$_{02}$, 3$_{22}$-2$_{21}$ and 3$_{21}$-2$_{20}$ transitions of H$_2$CO. The spectroscopic information of these transitions is shown in Table \ref{table:all_transitions} \citep{cdms2005}. The H$_2$CO line emission is very strong in all three sources with measured T$_{MB}>5K$ (top panels in Fig.~\ref{figure:all_spectra}). 

\subsection{APEX SEPIA observations}

The three sources were observed with the APEX SEPIA receiver \citep[Swedish-ESO PI receiver for APEX;][]{2012ITTST...2..208B}\footnote{Observations were done as part of Science Verification in July 2015 (project E-095.F-9808A), and in November-October 2015 and May-June 2016 within projects E-096.C-0484A and E097.C-0897A.}. We carried out single-pointing observations of HDCO and D$_2$CO using the position switching observing mode. The J2000 central coordinates used in our observations were [RA, Dec] = (05:30:48.0, +33:47:54) for AFGL5142, [RA, Dec] = (05:39:13.0, +35:45:51) for IRAS05358+3543, and [RA, Dec] = (18:00:30.5, -24:04:01) for G5.89-0.39.
The observed transitions of HDCO (at 185 GHz and 193 GHz) and D$_2$CO (at 174.4 GHz) are shown in Table \ref{table:all_transitions}. 
The pointing was checked every 60-90\,mins, and the typical system temperatures were 150\,K. The beam size at 183\,GHz was $\sim$34$\arcsec$ and a beam efficiency of 0.83 was used. The XFFTS spectrometer provided a velocity resolution of 0.059\,km\,s$^{-1}$ and 0.066\,km\,s$^{-1}$ for the HDCO (at 185-193\,GHz) and D$_2$CO transitions (at 174\,GHz), respectively. However, for the data analysis, we smoothed the spectra to a uniform velocity resolution of 0.5\,km/s. At this velocity resolution, we reached an rms of $\sim$ 0.01\,K with typical integration times between 32 and 48 minutes. The PWV was 0.3-1.5\,mm. The data were reduced and analyzed with the Gildas software \footnote{See http://www.iram.fr/IRAMFR/GILDAS.} \citep[version jul14a,][]{2005sf2a.conf..721P} and the derived parameters of the detected lines are reported in Appendix~\ref{app:lines}.

\section{Results}
In Fig.~\ref{figure:all_spectra}, we present the HDCO and D$_2$CO lines observed toward AFGL5142, IRAS~05358+3543, and G5.89-0.39. The H$_2$CO 3$_{0,3}$-2$_{0,2}$ data obtained with the IRAM 30m observation are also shown for comparison. While the HDCO lines (upper energy levels of 18.5-50.4\,K) are clearly detected in all sources with peak intensities $\geq$35 mK, the D$_2$CO 3$_{0,3}$-2$_{0,2}$ transition is only observed toward AFGL5142 and IRAS~05358+3543, i.e. at the earlier HMSC/HMPO stages. The 3$\sigma$ upper limit measured for the D$_2$CO 3$_{2,2}$-2$_{2,1}$ line toward G5.89-0.39 is $\leq$6\,mK. 

In Table \ref{table:fit}, we report the derived parameters of the observed lines of H$_2$CO, HDCO and D$_2$CO. For AFGL5142, the D$_2$CO 3$_{0,3}$-2$_{0,2}$ transition shows two velocity components at $-2.077$~km/s and $-3.658$~km/s (see Table~\ref{table:fit}). We thus fitted all the observed H$_2$CO, HDCO and D$_2$CO line emission using two Gaussian line profiles with fixed central velocities at $-2.077$~km/s and $-3.658$~km/s. These two velocity components are associated with the WC core at the HMSC stage (radial velocity $\sim -3.5$~km/s), and with the CC core at the HMPO stage \citep[at $\sim -2$~km/s;][]{2011A&A...525A.141B}. Both cores lie within the single-dish beam of our IRAM 30m and APEX observations (WC and CC are separated by 9$"$ while the IRAM 30m and APEX beam sizes are 11$"$ and 34$"$, respectively).

Since the H$_2$CO, HDCO and D$_2$CO lines toward G5.89-0.39 show a slight asymmetry in their line profiles, we have also fitted these data using two Gaussian line profiles centered at 8.5 and 11.5 km s$^{-1}$. These two components can also be seen in the emission from e.g. N$_2$H$^+$ or NH$_3$ \citep[see][]{2011A&A...529L...7F,2015A&A...575A..87F}. 

For IRAS~05358+3543, H$_2$CO, HDCO and D$_2$CO show only one velocity component peaking at $-$16\,km\,s$^{-1}$.

To estimate the molecular column densities of H$_2$CO, HDCO and D$_2$CO, we used the MADCUBAIJ software that assumes LTE conditions \citep[][]{martin2011, rivilla2016}. We adopted a source size of $6.\!^{\prime\prime}0$ as measured from NH$_3$(2,2) and following \citet{2015A&A...575A..87F}. This source size is consistent with that obtained in other high density tracers such as N$_2$H$^+$(1-0) and C$^{34}$S(7-6) \citep[see][]{2008ApJ...680.1271H,2011A&A...525A.141B}. We have attempted to derive the excitation temperature of the gas, T$_{\rm ex}$, using the four APEX HDCO transitions (Table~\ref{table:all_transitions}). Most of the observed J=3$\rightarrow$2 lines of H$_2$CO, HDCO and D$_2$CO likely trace gas with similar physical conditions given that their E$_{up}$ and A$_{ij}$ are similar. Hence, the determination of T$_{\rm ex}$ is highly uncertain due to the low number of available lines. We thus assumed a T$_{\rm ex}$ of 28~K, which is the average temperature obtained from NH$_3$ for IRAS\,05358+3545 and AFGL5142 \citep{2011A&A...529L...7F}\footnote{\citet{2011A&A...529L...7F} do not report any temperature for G5.89-0.39 and therefore we assume the same T$_{\rm ex}$ of 28\,K for this source.}. We note that the assumed value of T$_{\rm ex}$ can change the derived column densities, but not their ratios. Assuming a T$_{ex}$ value in the 20-40\,K range, the variation of the column density values compared to the T$_{ex}$=28\,K case is below 50\%. Lower and higher excitation temperatures can not fit the observed transitions properly. We have estimated the column densities with different source sizes also. With a source size of 2'' we can not fit the observed transitions properly due to optical depth effects. The estimated column densities decrease by factors of 2-3 for a source size of 10''. For AFGL5142, we obtained the H$_2$CO, HDCO and D$_2$CO column densities for the velocity components at $-2.077$~km/s (CC, in the HMPO stage) and at $-3.658$~km/s (WC in the HMSC phase). In Table \ref{table:deuteration}, we report the calculated column densities of H$_2$CO, HDCO and D$_2$CO and the deuteration fractions obtained from the column density ratios HDCO/H$_2$CO, D$_2$CO/H$_2$CO and D$_2$CO/HDCO. 

The H$_2^{13}$CO 3$_{1,3}$-2$_{1,2}$ transition was covered within our APEX SEPIA setup. We have estimated the H$_2$CO column densities from the H$_2^{13}$CO column density and using the $^{12}$C/$^{13}$C ratio derived by \citet{2005ApJ...634.1126M} as a function of Galactocentric distance. The H$_2$CO lines in AFGL5142 and G5.89-0.39 are not very optically thick since the H$_2$CO column densities calculated using the main isotopologue H$_2$CO and the $^{13}$C isotopologue H$_2^{13}$CO differ only by factors 0.4 and 3.6. For IRAS~05358+3543, however, the column densities derived from H$_2$CO and H$_2^{13}$CO differ by a factor of 9. In this case, we use the column density derived directly from the H$_2$CO lines and consider it as a lower limit for IRAS05358.

\begin{table*}[!h]
\small
\begin{tabular}{l c c c c c c c}
\hline
Name & Type	&	N(H$_2$CO)	&	N(HDCO)	&	N(D$_2$CO)	&	\small{N(HDCO)/N(H$_2$CO)} & \small{N(D$_2$CO)/N(H$_2$CO)} & \small{N(D$_2$CO)/N(HDCO)}	\\
   &     & (10$^{13}$~cm$^{-2}$) & (10$^{13}$~cm$^{-2}$) & (10$^{13}$~cm$^{-2}$) & & & \\
\hline
AFGL5142 &HMSC 	& 317 (10) & 11 (3) & 3.4 (2) & 0.035 (0.01) & 0.011 (0.01) & 0.31 (0.3) \\
IRAS05358+3543& HMSC/HMPO & $>$130 (2) & 13.0 (1.4) & 2.6 (1.7) & $<$0.10 (0.01) & $<$0.020 (0.01)& 0.20 (0.2) \\
AFGL5142 &HMPO   & 40 (2) & 14 (3) & 2.5 (1.6) & 0.34 (0.09) & 0.063 (0.04) & 0.18 (0.15) \\
G5.89-0.39  & UCHII    & 1740 (170)& 25 (2)& $<$2& 0.014 (0.002)& $<$0.001 & $<$0.08\\
\hline
\end{tabular}
\caption{Calculated column densities and ratios (deuteration fractions) for the three species. The errors for the calculated column density values are the formal fit uncertainties from the least square fits performed by MADCUBAIJ. Systematic uncertainties dominate over the formal fit uncertainties, see Section 3, paragraph 5. Calculated values based on the two velocity components of AFGL5142 - corresponding to the previously identified HMSC and HMPO objects by \citet{2011A&A...525A.141B} - are shown separately. \label{table:deuteration}}
\end{table*}

\section{Discussion}

As proposed by \citet{2011A&A...529L...7F, 2015A&A...575A..87F}, the different trends observed for D$_{frac}$(N$_2$H$^+$), D$_{frac}$(CH$_3$OH) and D$_{frac}$(NH$_3$) as a function of evolution in high-mass star-forming regions are likely due to the way deuteration occurs for the different species: in the gas phase for N$_2$H$^+$, on the grain surface for CH$_3$OH, and via a mixture of the two for NH$_3$. For H$_2$CO, D$_2$CO is expected to be a product of material processed on solid ices since its formation time-scales are longer than the depletion timescales in the pre-stellar phase \citep[see][]{2012ApJ...748L...3T}. Consequently, D$_2$CO should show an evolutionary trend similar to that derived for D$_{frac}$(CH$_3$OH), which peaks at the HMPO phase \citep{2015A&A...575A..87F}. If HDCO showed a similar behaviour to D$_2$CO, we could conclude that HDCO also forms mostly on dust grains. This hypothesis would be supported by the fact that HDCO is detected in hot cores and hot corinos while N$_2$D$^+$ (only formed in the gas phase) is not \citep[see e.g.][]{2005A&A...444..481F}.

\begin{figure}[!h]
  \centering
  \includegraphics[width=0.94\columnwidth]{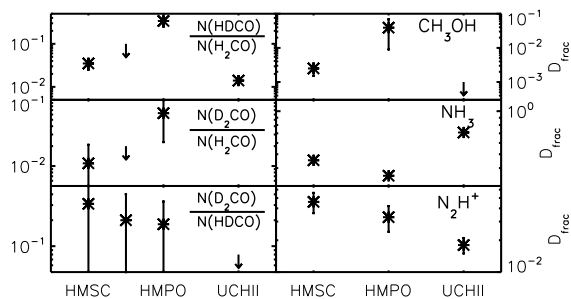}
  \caption{Comparison of the average deuterated fraction of H$_2$CO, HDCO, CH$_3$OH, NH$_3$ and N$_2$H$^+$. Upper limit for G5.89-0.39 is not shown on the left middle panel, the value is $<$0.001. Average values for CH$_3$OH, NH$_3$ and N$_2$H$^+$ based on \citet{2015A&A...575A..87F} are shown. \label{figure:dfrac_comparison}}
\end{figure}

Although our sample and statistics are limited, our results show a trend for the deuteration fractions HDCO/H$_2$CO and D$_2$CO/H$_2$CO to progressively increase from the HMSC to the HMPO stage by factors $\sim$3-10, and to subsequently decrease at the UCHII phase (by factors $\geq$20; see Fig. \ref{figure:dfrac_comparison}). This behaviour is similar to the trend observed for D$_{frac}$(CH$_3$OH), and since the measured HDCO/H$_2$CO are similar and D$_2$CO/H$_2$CO ratios are 4-21 times higher than those predicted by \citet{2007A&A...471..849R} for the production of HDCO and D$_2$CO via gas-phase reactions, surface chemistry likely plays an important role in the formation of HDCO and D$_2$CO.

\citet{2002P&SS...50.1125R} indeed proposed a grain surface formation scenario for these molecules in which H$_2$CO and, its deuterated forms, represent intermediate steps in the formation of CH$_3$OH and deuterated-CH$_3$OH via hydrogenation and D-addition reactions. The extent by which surface chemistry is responsible for the production of HDCO and D$_2$CO in these sources can be estimated by using the F parameter:
\begin{equation}
F=\frac{[HDCO/H_2CO]^2}{[D_2CO/H_2CO]}.
\label{Fpar}
\end{equation}

\noindent
From the HDCO/H$_2$CO and D$_2$CO/H$_2$CO ratios in Table~\ref{table:deuteration}, we find that F$\sim$0.1-0.5 for HMSCs, F$\sim$1.8 for HMPOs and F$\geq$0.2 for UCHIIs. As explained by \citet[][]{2002P&SS...50.1125R}, low F values are consistent with grain surface chemistry. However, F values in the range 1.6-2.3 could be explained also by gas-phase chemistry. Therefore, at the HMPO stage we cannot rule out a contribution to the HDCO/D$_2$CO production via the gas phase neutral-neutral reactions CH$_2$D + O $\rightarrow$ HDCO + H and CHD$_2$ + O $\rightarrow$ D$_2$CO + H. Note that these reactions occur in the forward direction at temperatures $\sim$ 30-50\,K due to the exothermicity of the formation reactions of CH$_2$D and CHD$_2$ \citep{1990ApJ...362L..29T}.  

Table~\ref{table:deuteration} also shows that the D$_2$CO/HDCO ratio remains constant at $\sim$0.2 within the uncertainties regardless of the source or evolutionary stage. As for the low-mass regime, the ratio HDCO/D$_2$CO lies well below the statistical value given as D-species/D$_2$-species=4$\times$(D-species/H-species)$^{-1}$, consistent with the grain surface formation scenario \citep[see][]{2014prpl.conf..859C}. 

If we compare the HDCO/H$_2$CO and D$_2$CO/H$_2$CO ratios from Table~\ref{table:deuteration} with those collected by \citet{2014prpl.conf..859C} for low-mass star-forming regions (see their Figure 5), we find that the ratio $\frac{HDCO/H_2CO}{D_2CO/H_2CO}$$\sim$3-5 in the high-mass regime while the same ratio is close to 1 in low-mass star-forming regions. This may be a consequence of the shorter time-scales available in the high-mass regime for the formation of D$_2$CO in the ices \citep{2012ApJ...748L...3T, 2002P&SS...50.1125R}. Indeed, in the scheme of Rodgers \& Charnley (2002), D$_2$CO is formed after HDCO through two hydrogen-deuterium exchanges experienced by H$_2$CO in the ices. The higher $\frac{HDCO/H_2CO}{D_2CO/H_2CO}$ ratio in high-mass star-forming regions therefore suggests that HDCO may not have had enough time to get converted into D$_2$CO as a consequence of the faster evolution of high-mass protostars as compared to their low-mass counterparts.

\section{Summary}
We have observed HDCO and D$_2$CO toward a sample of high-mass star-forming regions at different evolutionary stages. HDCO transitions were detected for all of them, while the D$_2$CO line was detected only for the earlier HMSC and HMPO stages. Our results point toward the idea that H$_2$CO, and its deuterated species, form mostly on grain surfaces although some gas-phase contribution is expected at the warm HMPO stage. Interferometric observations are needed to separate the HDCO and D$_2$CO emission originating from the small and dense cores and to disentangle their origin in high-mass star-forming regions.

\begin{acknowledgements}
We would like to thank the SEPIA team and APEX team for the successful commissioning of the instrument and for the observations. 
This work was partly supported by the Gothenburg Centre of Advanced Studies in Science and Technology through the program {\it Origins of habitable planets} and by the Italian Ministero dell'Istruzione, Universit\`a e Ricerca through the grant Progetti Premiali 2012 -- iALMA (CUP C52I13000140001) and by NAOJ ALMA Scientific Research Grant Number 2016-03B. K.W. acknowledges the support from Deutsche Forschungsgemeinschaft (DFG) grant WA3628-1/1 through priority programme 1573 ('Physics of the Interstellar Medium''). I.J.-S. acknowledges the financial support received from the STFC through an Ernest Rutherford Fellowship (proposal number ST/L004801/1). L.V.T. and S.Z. acknowledge the support by the OTKA grants NN-111016 and K101393.
\end{acknowledgements}

\bibliographystyle{aa}
\bibliography{references}

\appendix
\section{Observed H$_2$CO, HDCO, and D$_2$CO lines}
\label{app:lines}

In Table~\ref{table:all_transitions} we report the parameters of the observed lines for H$_2$CO, HDCO, and D$_2$CO. The observing setup \citep[and the observations of][]{2011A&A...529L...7F,2015A&A...575A..87F} also covered other, higher excitation lines of the same molecules, but these were not detected (and are not expected to be detectable under the expected excitation conditions).

\begin{figure*}[!h]
  \centering
  \includegraphics[width=0.6\columnwidth]{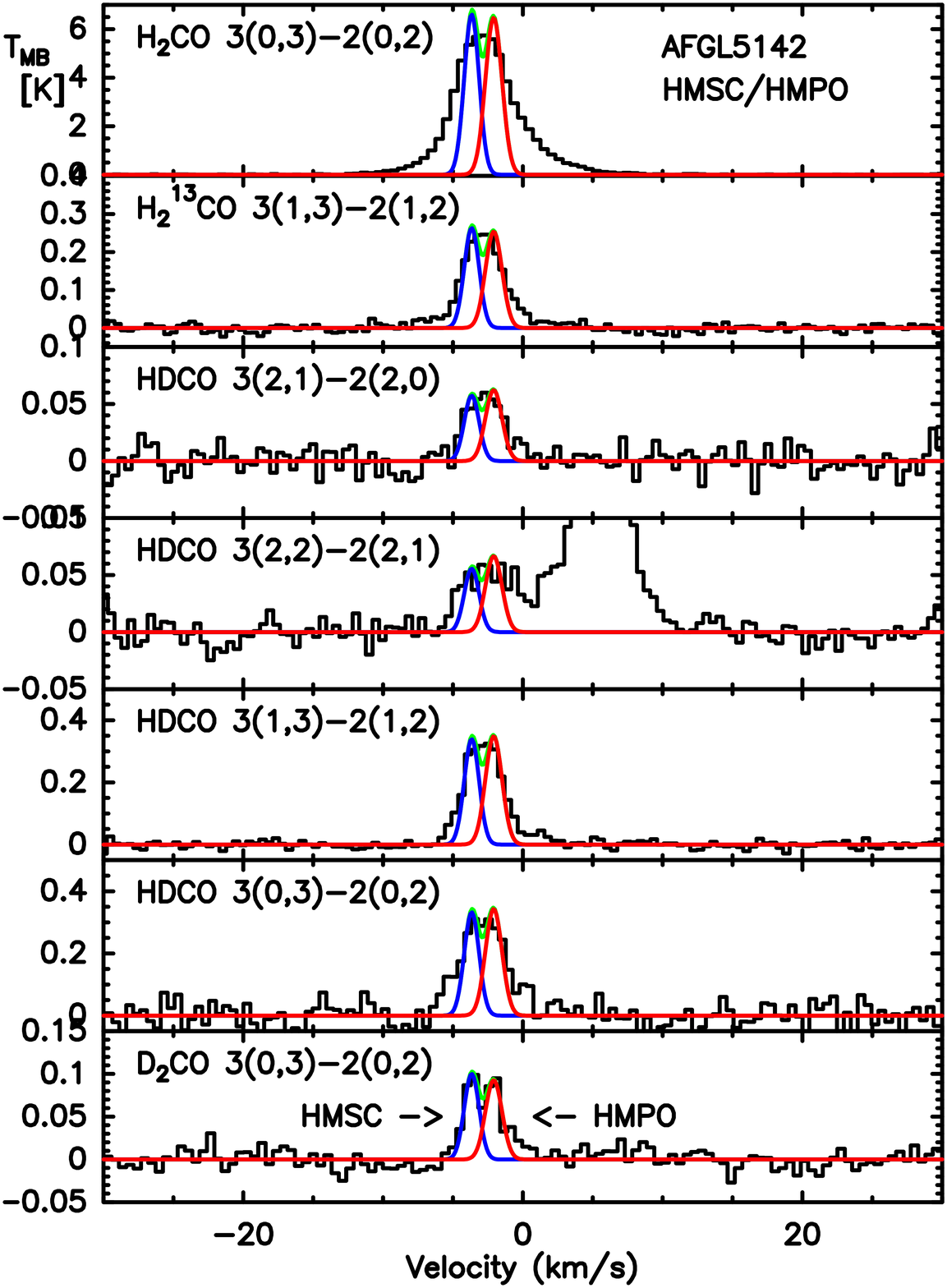}
  \includegraphics[width=0.6\columnwidth]{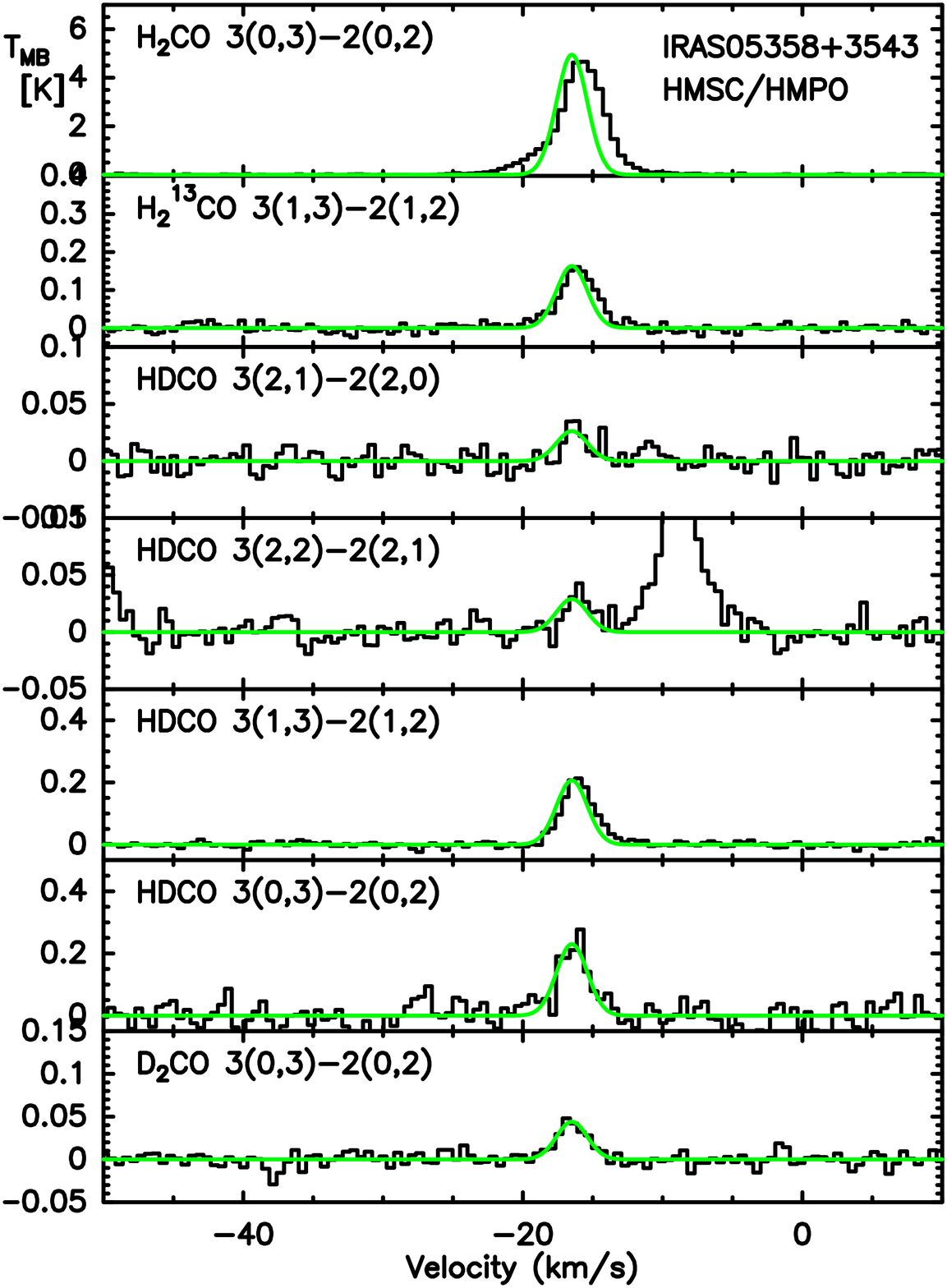}
  \includegraphics[width=0.6\columnwidth]{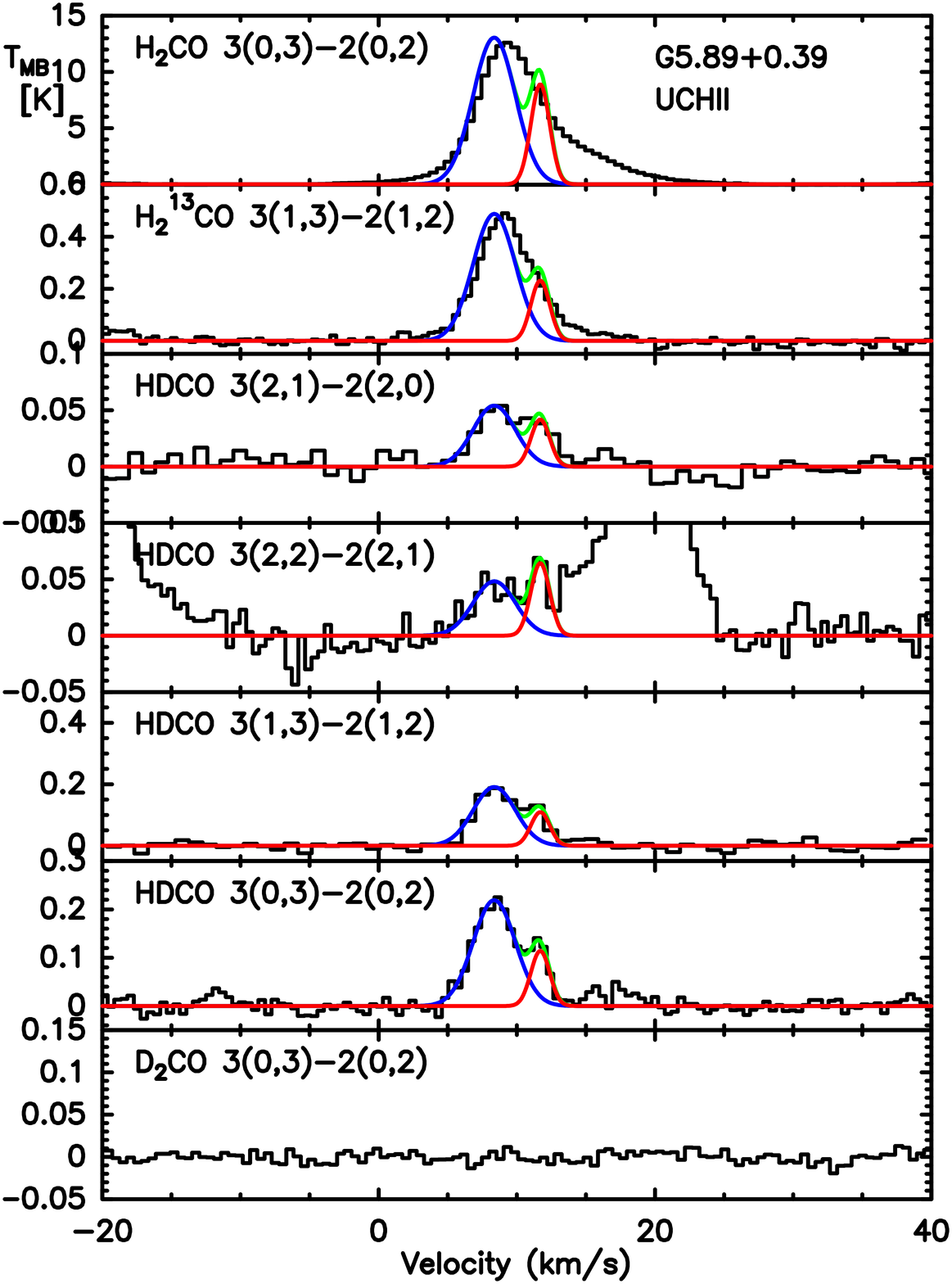}
  \caption{From left to right: Spectra of H$_2$CO 3$_{0,3}$-2$_{0,2}$, H$_2^{13}$CO 3$_{1,3}$-2$_{1,2}$, HDCO 3$_{2,1}$-2$_{2,0}$, 3$_{2,2}$-2$_{2,1}$, 3$_{1,3}$-2$_{1,2}$, 3$_{0,3}$-2$_{0,2}$ and D$_2$CO 3$_{0,3}$-2$_{0,2}$ observed towards AFGL5142, IRAS~05358+3543, and G5.89-0.39 using the IRAM 30m telescope \citep[][]{2011A&A...529L...7F,2015A&A...575A..87F} and the APEX SEPIA Band 5 receiver (this work). Red and blue lines indicate the fitted different velocity components for AFGL5142 and G5.89-0.39 with fixed velocity and width. The C$^{13}$CS N=15-14, J=15-14 transition is seen on the fourth panels from the top. \label{figure:all_spectra}}
\end{figure*}

In Fig.~\ref{figure:all_spectra}, we present the HDCO and D$_2$CO lines observed toward AFGL5142, IRAS~05358+3543, and G5.89-0.39. The H$_2$CO 3$_{0,3}$-2$_{0,2}$ data obtained with the IRAM 30m observation are also shown for comparison. While the HDCO lines (upper energy levels of 18.5-50.4\,K) are clearly detected in all sources with peak intensities $\geq$35 mK, the D$_2$CO 3$_{0,3}$-2$_{0,2}$ transition is only observed toward AFGL5142 and IRAS~05358+3543, i.e. at the earlier HMSC/HMPO stages. The 3$\sigma$ upper limit measured for the D$_2$CO 3$_{2,2}$-2$_{2,1}$ line toward G5.89-0.39 is $\leq$6\,mK. 

In Table~\ref{table:fit} we report the parameters derived fitting single or multiple (in the case of AFGL5142 and G5.89-0.39) Gaussian profiles to the detected lines.  For the HDCO, H$_2^{13}$CO and H$_2$CO spectra of AFGL5142 and IRAS05358+3543, we fitted the v$_{LSR}$ and FWHM values based on the D$_2$CO velocity components. For G5.89-0.39, we derived the velocity and width of the two components based on the HDCO 3$_{0,3}\rightarrow$2$_{0,2}$ line. We do not use directly these fits for the calculation of the physical parameters. We use the velocities and linewidths as initial parameters for the MADCUBAIJ modeling.

\begin{table}[!h]
\begin{tabular}{l c c c c c}
\hline
Species	& Transition &	Frequency	&	E$_{up}$	&	A$_{ij}$	& Telescope \\
	&	&	[GHz]	&	[K]	&	[s$^{-1}$]	& \\
\hline
D$_2$CO 	&	3$_{0,3}$$\rightarrow$2$_{0,2}$ 	& 174.413	&	 16.8	&	1.44e-04 & APEX \\
HDCO 	&	3$_{1,3}$$\rightarrow$2$_{1,2}$ 	& 185.307	&		25.8	&	1.53e-04 & APEX	\\
HDCO 	&	3$_{0,3}$$\rightarrow$2$_{0,2}$ 	& 192.893	&		18.5	&	1.94e-04 & APEX \\
HDCO 	&	3$_{2,2}$$\rightarrow$2$_{2,1}$ 	& 193.392	&		50.4	&	1.09e-04 & APEX	\\
HDCO 	&	3$_{2,1}$$\rightarrow$2$_{2,0}$ 	& 193.907	&		50.4	&	1.10e-04 & APEX	\\
H$_2^{13}$CO 	&	3$_{1,3}$$\rightarrow$2$_{1,2}$ 	& 206.131	&		31.6	&	2.11e-04 & APEX	\\
\hline
H$_2$CO 	&	9$_{1,8}$$\rightarrow$9$_{1,9}$ 	& 216.569	&		174.0	&	7.22e-06 & IRAM	\\
H$_2$CO 	&	3$_{0,3}$$\rightarrow$2$_{0,2}$ 	& 218.222	&		21.0	&	2.82e-04 & IRAM	\\
H$_2$CO 	&	3$_{2,2}$$\rightarrow$2$_{2,1}$ 	& 218.476	&		68.1	&	1.57e-04 & IRAM	\\
H$_2$CO 	&	3$_{2,1}$$\rightarrow$2$_{2,0}$ 	& 218.760	&		68.1	&	1.58e-04 & IRAM	\\
\hline
\end{tabular}
\caption{Transitions of H$_2$CO, H$_2^{13}$CO, HDCO and D$_2$CO observed with APEX SEPIA and the IRAM-30m telescope. \label{table:all_transitions}}
\end{table}

\begin{table*}[!h]
\begin{tabular}{l c r r r r}
\hline
Line & Area & v$_{LSR}$ & FWHM & T$_{\rm MB}$ \\
 & [K km/s] & [km/s] & [km/s] & [K]	\\
\hline
\multicolumn{5}{c}{\textbf{AFGL5142}} \\
D$_2$CO 3$_{0,3}$$\rightarrow$2$_{0,2}$ & 0.13 (0.01) & \textbf{-3.66 (0.06)}  & 1.3 (0.2) & 0.10 (0.01) \\
& 0.14 (0.01) & \textbf{-2.08 (0.05)} &	1.4	(0.3) & 0.09 (0.01)	\\
HDCO 3$_{0,3}$$\rightarrow$2$_{0,2}$ & 0.45 (0.01) 	& -3.66	&	1.3	&	 0.34 (0.01) \\
 &  0.52 (0.01) & -2.08 & 1.4 & 0.35 (0.01) \\
HDCO 3$_{1,3}$$\rightarrow$2$_{1,2}$ & 0.44 (0.04) 	& -3.66	& 1.3	&	 0.33 (0.04) \\
 &	 0.51 (0.04) 	&	-2.08	&	1.4	&	 0.34 (0.04) \\
HDCO 3$_{2,2}$$\rightarrow$2$_{2,1}$&	 0.07 (0.06) 	&	-3.66	&	1.3	&	"	0.06 (0.01)	\\
&	0.10 (0.06)	&	-2.08	&	1.4	&	0.07 (0.01) \\
HDCO 3$_{2,1}$$\rightarrow$2$_{2,0}$&	 0.07 (0.01) 	&	-3.66	&	1.3	&	 0.06 (0.01) \\
&	 0.09 (0.01) 	&	-2.08	&	1.4	&	 0.06 (0.01) \\
H$_2^{13}$CO 3$_{1,2}$$\rightarrow$2$_{1,2}$	&	0.35 (0.01)	&	-3.66	&	1.3	&	0.26 (0.01) \\
	&	0.38 (0.01)	&	-2.08	&	1.4	&	0.25 (0.01) \\
H$_2$CO 3$_{0,3}$$\rightarrow$2$_{0,2}$	&	 8.81 (0.48) 	&	-3.66	&	1.3	&	 6.60 (0.48) \\
&	 9.62 (0.51) 	&	-2.08	&	1.4	&	 6.47 (0.48) \\
\multicolumn{5}{c}{\textbf{IRAS05358+3543}}	\\
D$_2$CO 3$_{0,3}$$\rightarrow$2$_{0,2}$	&	0.12	(0.01)	&	 -16.49 (0.15) 	&	 2.6 (0.4) 	&	0.04 (0.01)	\\
HDCO 3$_{0,3}$$\rightarrow$2$_{0,2}$&	 0.57 (0.01) 	&	-16.49	&	2.6	&	 0.21 (0.01) \\
HDCO 3$_{1,3}$$\rightarrow$2$_{1,2}$&	 0.64 (0.05) 	&	-16.49	&	2.6	&	 0.23 (0.04) \\
HDCO 3$_{2,2}$$\rightarrow$2$_{2,1}$	&	 0.08 (0.04) 	&	-16.49	&	2.6	&	 0.03 (0.01) \\
HDCO 3$_{2,1}$$\rightarrow$2$_{2,0}$&	 0.07 (0.01) 	&	-16.49	&	2.6	&	 0.03 (0.01) \\
H$_2^{13}$CO 3$_{1,2}$$\rightarrow$2$_{1,2}$	&	0.45 (0.01)	&	-16.49	&	2.6	&	0.16 (0.01) \\
H$_2$CO 3$_{0,3}$$\rightarrow$2$_{0,2}$	&	 13.70 (0.14) 	&	-16.49	&	2.6	&	 4.95 (0.09) \\
\multicolumn{5}{c}{\textbf{G5.89-0.39}}	\\
D$_2$CO 3$_{0,3}$$\rightarrow$2$_{0,2}$	 &	 $\leq$0.03 	&	 $\ldots$ 	&	 $\ldots$ 	&	 $\leq$0.02 \\
HDCO 3$_{0,3}$$\rightarrow$2$_{0,2}$ 	&	  0.84(0.03) 	&	  8.37(0.08)  	&	 3.60 (0.18) 	&	    0.22 (0.01) \\
&	  0.19(0.02) 	&	 11.70 (0.08)    	&	 1.60 (0.21) 	&	    0.11 (0.01) \\
HDCO 3$_{1,3}$$\rightarrow$2$_{1,2}$ 	&	  0.73(0.03) 	&	8.37	&	3.6	&	    0.19 (0.01) \\
&	  0.19(0.02) 	&	11.7	&	1.6	&	    0.11 (0.01) \\
HDCO 3$_{2,2}$$\rightarrow$2$_{2,1}$ 	&	  0.18(0.11)   	&	8.37	&	3.6	&	    0.05 (0.01) \\
&	  0.11(0.08)   	&	11.7	&	1.6	&	    0.06 (0.01) \\
HDCO 3$_{2,1}$$\rightarrow$2$_{2,0}$ 	&	  0.21(0.03) 	&	8.37	&	3.6	&	    0.05 (0.01) \\
&	  0.07(0.01) 	&	11.7	&	1.6	&	    0.04 (0.01) \\
H$_2^{13}$CO 3$_{1,2}$$\rightarrow$2$_{1,2}$	&	1.87 (0.03)	&	8.37	&	3.6	&	0.48 (0.02) \\
	&	0.39 (0.02)	&	11.7	&	1.6	&	0.23 (0.02)\\
H$_2$CO 3$_{0,3}$$\rightarrow$2$_{0,2}$ 	&	  50.03(1.61) 	&	8.37	&	3.6	&	     13.05 (0.93) \\
&	  15.16(1.07) 	&	11.7	&	1.6	&	     8.90 (0.93) \\
\hline
\end{tabular}
\caption{Fitted parameters of the observed H$_2$CO, H$_2^{13}$CO, HDCO and D$_2$CO lines. Errors in peak intensity correspond to the 1\,$\sigma$ rms level in the spectra. For the HDCO, H$_2^{13}$CO and H$_2$CO spectra of AFGL5142 and IRAS05358+3543, we fitted the v$_{LSR}$ and FWHM values based on the D$_2$CO velocity components. For G5.89-0.39, we derived the velocity and width of the two components based on the HDCO 3$_{0,3}\rightarrow$2$_{0,2}$ line. \label{table:fit}}
\end{table*}

\Online

%
%

\end{document}